\documentclass[prl,twocolumn,superscriptaddress,amsmath,amssymb]{revtex4}

\usepackage{array}
\usepackage{graphicx} %\graphicspath{{../figures/}}
\usepackage{color}
\usepackage{bm}

\def\beq{\begin{equation}}
\def\eeq{\end{equation}}
\def\bea{\begin{eqnarray}}
\def\eea{\end{eqnarray}}
\def\fatR{{\bf R}}

\def\fatM{{\bf M}}

\begin{document}

\title{Fast and Accurate Modeling of Molecular Atomization Energies with Machine Learning}

\author{Matthias Rupp}
\affiliation{Machine Learning Group, Technical University of Berlin, Franklinstr 28/29, 10587 Berlin, Germany}
\affiliation{Institute of Pure and Applied Mathematics, University of California Los Angeles, Los Angeles, CA 90095, USA} 
\author{Alexandre Tkatchenko}
\affiliation{Fritz-Haber-Institut der Max-Planck-Gesellschaft, 14195 Berlin, Germany}
\affiliation{Institute of Pure and Applied Mathematics, University of California Los Angeles, Los Angeles, CA 90095, USA} 
\author{Klaus-Robert M\"uller}
\affiliation{Machine Learning Group, Technical University of Berlin, Franklinstr 28/29, 10587 Berlin, Germany}
\affiliation{Institute of Pure and Applied Mathematics, University of California Los Angeles, Los Angeles, CA 90095, USA} 
\author{O. Anatole von Lilienfeld}
\email{anatole@alcf.anl.gov}
\affiliation{Argonne Leadership Computing Facility, Argonne National Laboratory, Argonne, Illinois 60439, USA}
\affiliation{Institute of Pure and Applied Mathematics, University of California Los Angeles, Los Angeles, CA 90095, USA} 

\date{\today}

\begin{abstract}
We introduce a machine learning model to predict atomization energies of a diverse set of organic molecules, based on nuclear charges and atomic positions only.
The problem of solving the molecular Schr\"odinger equation is mapped onto a non-linear statistical regression problem of reduced complexity.
Regression models are trained on and compared to atomization energies computed with hybrid density-functional theory.
Cross-validation over more than seven thousand small organic molecules yields a mean absolute error of $\sim$10~kcal/mol.
Applicability is demonstrated for the prediction of molecular atomization potential energy curves.  
\end{abstract}

\maketitle

%{\em Introduction---}
Solving the Schr\"odinger equation (SE), $H\Psi = E\Psi$, for assemblies of atoms is a fundamental problem in quantum mechanics.
Alas, solutions that are exact up to numerical precision are intractable for all but the smallest systems with very few atoms. 
Hierarchies of approximations have evolved, usually trading accuracy for computational efficiency~\cite{Encyclopedia}.
%In this Letter, we answer the fundamental question: Provided with a finite number of solved examples, 
%can we estimate energies of new systems without the need for 
%mean-field approximations and self-consistent field (SCF) procedures? 
Conventionally, the external potential, defined by a set of nuclear charges $\{Z_I\}$ and atomic positions $\{\fatR_I\}$,
uniquely determines the Hamiltonian $H$ of {\em any} system,
%for a non-degenerate ground state of a system with a given charge, 
and thereby the potential energy by optimizing $\Psi$,\cite{HK}
$H(\{Z_I, \fatR_I\}) \stackrel{\rm \Psi}{\longmapsto} E$.
For a diverse set of organic molecules, we show that one can use machine learning (ML) instead, 
$\{Z_I, \fatR_I\} \stackrel{\rm ML}{\longmapsto} E$.
Thus, we circumvent the task of explicitly solving the SE
by training once a machine on a finite subset of known solutions. 
Since many interesting questions in physics require to repeatedly solve the SE,
the highly competitive performance of our ML approach may pave the way to large 
scale exploration of molecular energies in chemical compound space~\cite{ChemicalSpace,anatole-jcp2006-2}. 

ML techniques have recently been used with success to map the problem of 
solving complex physical differential equations to statistical models. 
Successful attempts include solving %Navier-Stokes 
Fokker-Planck stochastic differential equations \cite{cklmn2008},
parameterizing interatomic force fields for fixed chemical composition~\cite{bpkc2010,MachineLearningWaterPotential_Handley2008}, 
and the discovery of novel ternary oxides for batteries~\cite{MachineLearningHautierCeder2010}.
Motivated by these, and other related 
efforts~\cite{PotentialEnergyFit_Bowman2003,Neuralnetworks_Scheffler2004,Neuralnetworks_BehlerParrinello2007,Neuralnetworks_Behler2008}, 
we develop a non-linear regression ML model for computing molecular atomization energies in 
chemical compound space~\cite{ChemicalSpace}.
Our model is based on a measure of distance in compound space that accounts for both stoichiometry and configurational variation.
After training, energies are predicted for new (out-of-sample) molecular systems, differing in composition and geometry, 
at negligible computational cost, {\em i.e.} milli seconds instead of hours on a conventional CPU. 
While the model is trained and tested using atomization energies calculated at the 
hybrid density-functional theory (DFT) level~\cite{HK,KS,becke3}, 
any other training set or level of theory could be used as a starting point for subsequent ML training.
%Cross-validation for our largest training set yields a mean absolute error of 8.4~kcal/mol.
Cross-validation on 7165 molecules yields a mean absolute error of 9.9~kcal/mol,
which is an order of magnitude more accurate than counting bonds or semi-empirical quantum chemistry.
%what can we do now:

%{\em Methodology---}
We use the GDB data base, a library of nearly one billion organic molecules that
are stable and synthetically accessible according to organic chemistry rules~\cite{ReymondChemicalUniverse3}.
While potentially applicable to any stoichiometry, as a proof of principle we restrict ourselves to small organic molecules.
Specifically, we define a controlled test-bed consisting of all 7165 organic molecules from the GDB data base
%\cite{ReymondChemicalUniverse,ReymondChemicalUniverse2}
with up to seven ``heavy'' atoms that contain C, N, O, or S, being saturated with hydrogen atoms. 
Atomization energies range from -800 to -2000 kcal/mol.  
Structural features include a rich variety of chemistry such as double, 
and triple-bonds; (hetero)cycles, carboxy, cyanide, amide, alcohol, and epoxy-groups.
For each of the many stoichiometries, many constitutional (differing chemical bonds) 
but no conformational isomers are part of this data base. 
Based on the string representation of molecules in the data base, %(SMILES[CITE]), 
we generated Cartesian geometries with OpenBabel~\cite{OpenBabel}. 
Thereafter, the PBE0 approximation to hybrid DFT~\cite{PBE0,PBE01} in converged numerical basis, 
as implemented in the {\tt FHI-aims} code~\cite{aims} (tight settings/tier2 basis set), was used to compute reference atomization energies for training.
%Open shell calculations of respective most stable electronic spin states were carried out for the corresponding free atom energies.
Our choice of the PBE0 hybrid functional is motivated by small errors ($<$ 5~kcal/mol) for thermo-chemistry 
data that includes molecular atomization energies~\cite{DFT4G3_Truhlar2003}.

One of the most important ingredients for ML is the choice of an appropriate data representation
that reflects prior knowledge of the application domain, {\em i.e.} a model of the underlying physics. 
A variety of such ``descriptors'' are used by statistical methods for chem- and bio-informatics 
applications~\cite{SchneiderReview2010,WienerDescriptors,SignatureFaulon2003}.
For modeling atomization energies, we use the same molecular information that enters the Hamiltonian for an 
electronic structure calculation, namely the set of Cartesian coordinates, $\{\fatR_I\}$, and nuclear charges, $\{Z_I\}$.
Our representation consists of atomic energies, 
and the inter-nuclear Coulomb repulsion operator, 
%\bea
%\hat{V}_{C} & =& \frac{1}{2} \sum_{I \ne J} \frac{Z_I Z_J}{|\fatR_I-\fatR_J|},
%\label{eq:coulomb}
%\eea
%where $Z_I$ corresponds to the nuclear charge of atom $I$ at $\fatR_I$.
Specifically, we represent {\em any} molecule by a ``Coulomb'' matrix $\fatM$,
\bea
M_{IJ}  =
\begin{cases}
  0.5 Z_I^{2.4} & \forall \;\; I = J,\\
   \frac{Z_IZ_J}{|\fatR_I - \fatR_J|} & \forall \;\; I \ne J.
\end{cases}
\label{eq:matrix}
\eea
Here, off-diagonal elements correspond to the Coulomb repulsion between atoms $I$ and $J$,
while diagonal elements encode a polynomial fit of atomic energies to nuclear charge.

Using ML we attempt to construct a non-linear map between molecular characteristics and atomization energies.   
This requires a measure of molecular (dis)similarity that is invariant with respect to translations, rotations, and the index ordering of atoms.
To this end, we measure the distance between two molecules by the Euclidean norm of their
diagonalized Coulomb matrices:
$d(\fatM,\fatM') = d({\bm \epsilon},{\bm \epsilon}') = \sqrt{\sum_{I} |\epsilon_{I} - \epsilon'_{I}|^2}$,
where ${\bm \epsilon}$ are the eigenvalues of $\fatM$ in order of decreasing absolute value.
For matrices that differ in dimensionality, ${\bm \epsilon}$ of the smaller system is extended by zeros. 
% MR: Extending epsilon is what was actually done, and is equivalent to extending M before diagonalization.
Note that by representing chemical compound space in this way, 
(i) any system is uniquely encoded because stoichiometry as well as atomic configuration are explicitly accounted for,
(ii) symmetrically equivalent atoms contribute equally, 
(iii) the diagonalized $\fatM$ is invariant with respect to atomic permutations, translations, and rotations, and
(iv) the distance is continuous with respect to small variations in inter-atomic distances or nuclear charges.
As discussed in Ref.~\cite{Neuralnetworks_Behler2011}, these are all crucial criteria 
for representing atomistic systems within statistical models.

\begin{figure}[htbp]
\centering
\includegraphics[scale=0.6, angle=0]{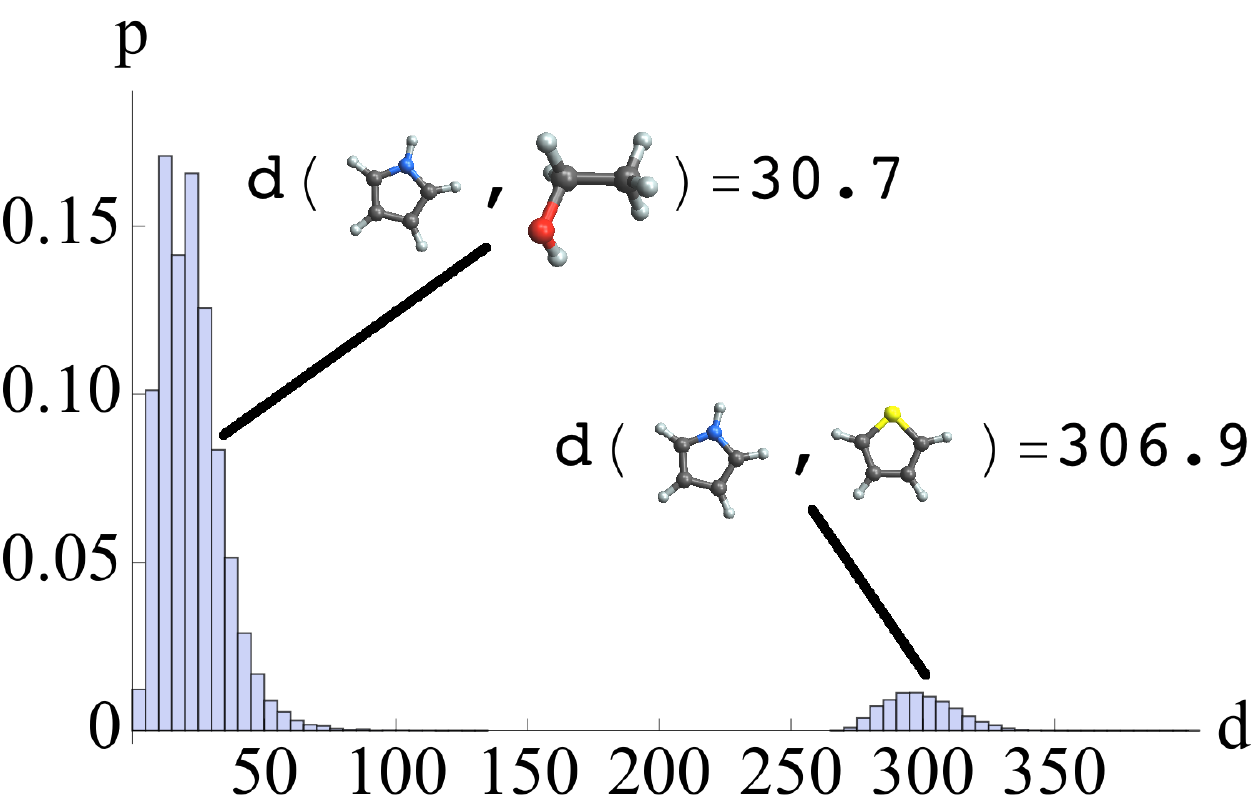}
\includegraphics[scale=0.3, angle=270]{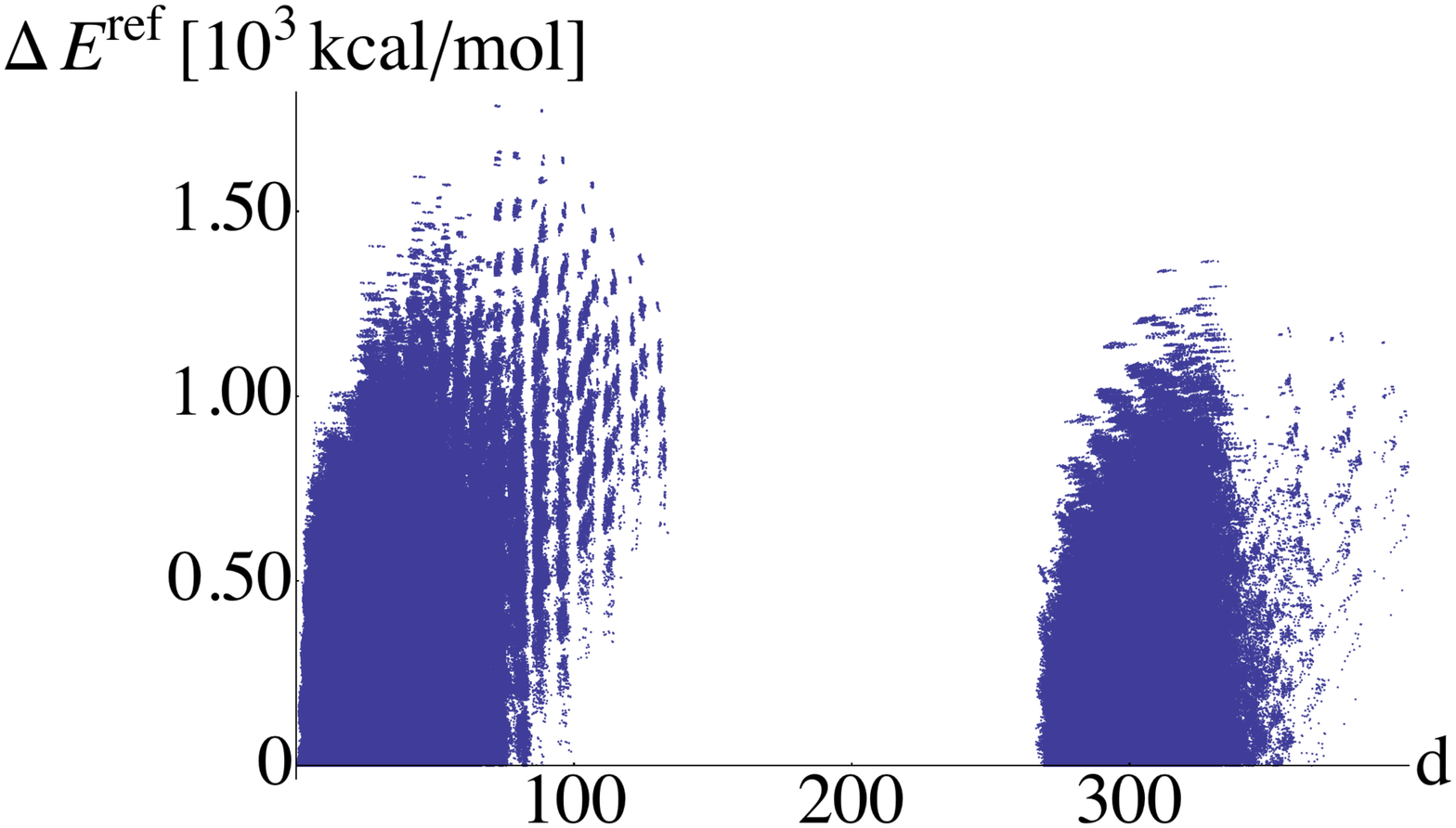} % was 0.3 270
\caption{(Color online) Top:
Distribution of distances, $d(\fatM,\fatM')$, for all molecular pairs occurring  
in the first 7165 small organic molecules from the GDB data base~\cite{ReymondChemicalUniverse3}.
The inset exemplifies two distances, pyrrol/ethanol and pyrrol/thiophene (N: Blue, O: Red, S: Yellow, C: Black, H: White). 
Bottom: Absolute differences in atomization energies between $\fatM$ and $\fatM'$ as a function of $d(\fatM,\fatM')$.
}
\label{fig:distance}
\end{figure}

In Fig.~\ref{fig:distance}, relative atomization energies, as a function of $d(\fatM,\fatM')$, and a histogram of distances are shown for all pairs of molecules in our data set. 
The inset exemplifies the distances between three molecular species, pyrrol, thiophene, and ethanol: 
Within our measure of similarity the nitrogen containing aromatic heterocycle pyrrol is $\sim$10 times farther
away from its sulfur containing analogue, thiophene, than from ethanol. 
This is due to the large difference in nuclear charges between atoms from different rows in the periodic table.

%\begin{figure*}[ht]
%\end{figure*}

Within our ML model~\cite{MLbook,htf2009,MueMikRaeTsuSch01}, 
%the Coulomb matrices $\fatM$ of all molecules in a training set,
the energy of a molecule $\fatM$ is a sum over weighted Gaussians, 
%This requires a kernel, {\em i.e.} a function $k$ of any two $\fatM$ and $\fatM'$.
%we employ a standard Gaussian kernel, 
\bea
E^{est}(\fatM) & = & \sum_{i=1}^N \alpha_i \exp \biggl[-\frac{1}{2 \sigma^2} d(\fatM,\fatM_i)^2 \biggr],
%+ b
%k(\fatM,\fatM') & = & e^{- \frac{1}{2 \sigma^2} d(\fatM,\fatM')^2 }.
%k(\fatM,\fatM') & = & \mathrm{exp} \biggl( -\frac{d(\fatM,\fatM')^2}{2 \sigma^2} \biggr) ,
\label{eq:energy}
\eea
where $i$ runs over all molecules $\fatM_i$ in the training set. 
Regression coefficients $\{\alpha_i\}$ and length-scale parameter $\sigma$ 
are obtained from training on $\{\fatM_i, E^{ref}_i\}$. 
Note that each training molecule $i$ contributes to the energy not only
according to its distance, but also according to its specific weight $\alpha_i$.
%This permits energy classification in the chemical space employed for training. %% MR: What does this mean? Please rephrase.
The $\{E^{ref}_i\}$ were computed at PBE0 DFT level of theory.

%are nonlinearly projected into a high-dimensional inner product space, 
%where inference (here, ridge regression) is performed implicitly,
% AT: We need to say what KRR is in layman terms, and why we use KRR and not something else.
%a nonlinear regularized regression model, %Model complexity is controlled by limiting 
To determine $\{ \alpha_i\}$, we used kernel ridge regression~\cite{htf2009}.
This regularized model limits the norm of regression coefficients, $\{ \alpha_i \}$,
thereby ensuring the transferability of the model to new compounds. 
For given length-scale $\sigma$ and regularization parameter $\lambda$, 
the explicit solution to the minimization problem,
\bea
\underset{{\bm \alpha}}{\rm min} && \sum_i \bigl( E^{est}(\fatM_i) - E^{ref}_i \bigr)^2 + \lambda \sum_i \alpha_i^2 \label{eq:krr}
\eea
is given by ${\bm \alpha}  =  ({\bf K} + \lambda {\bf I})^{-1} {\bf E}^{ref}$,
$K_{ij}$ = $\exp[-d(\fatM_i,\fatM_j)^2/(2\sigma^2)]$ 
being the kernel matrix of all training molecules, 
and ${\mathbf{I}}$ denoting the identity matrix. 

We used stratified~\cite{stratification} five-fold 
cross-validation~\cite{htf2009,MueMikRaeTsuSch01} for model selection and to estimate performance. 
Parameters $\lambda$ and $\sigma$ were determined in an inner loop of five-fold cross-validation
using a logarithmically scaling grid. 
This procedure is routinely applied in machine learning and statistics to 
avoid over-fitting and overly optimistic error estimates. 
%Note that all reported prediction errors in this work are out-of-sample predictions on compounds that were not used during training. 
%Those ML models that were employed for predictions of novel molecules
%use $\lambda$ and $\sigma$ averaged over cross-validation.
%is avoided by respecting the above cross-validation protocol, 
%
%we solved the regularized least square optimization problem denoted in Eqs.~(\ref{eq:krr}).
%(iii) Once the hyperparameters $(\lambda`,\sigma`)$ have
%been fixed, the resulting predictive model can be
%queried by an out-of-sample compound $\fatM$ to yield an estimate of the formation energy $E^{est}(\fatM)$
%from Eq.~(\ref{eq:predictor}) using the parameters ${\bf \alpha},b$ from the regression. 
%optimization at $(\lambda`,\sigma`)$ in Eq.~(\ref{eq:krr}). 
%(ii) To statistically estimate performance, we used 
%stratified five-fold cross-validation~\cite{htf2009,MueMikRaeTsuSch01}. 
%For each cross-validation fold we determined the hyperparameters $\lambda$ (responsible for adapting model
%stiffness) and $\sigma$ (the kernel scale) using a second, inner loop of 5-fold cross-validation. 
%Finally, the best model (smallest cross-validated mean absolute error) with parameters $(\lambda`,\sigma`)$
%%is used in the outer loop of cross-validation to predict the test fold.
%

\begin{figure}[ht]
\centering
\includegraphics[scale=0.6, angle=0]{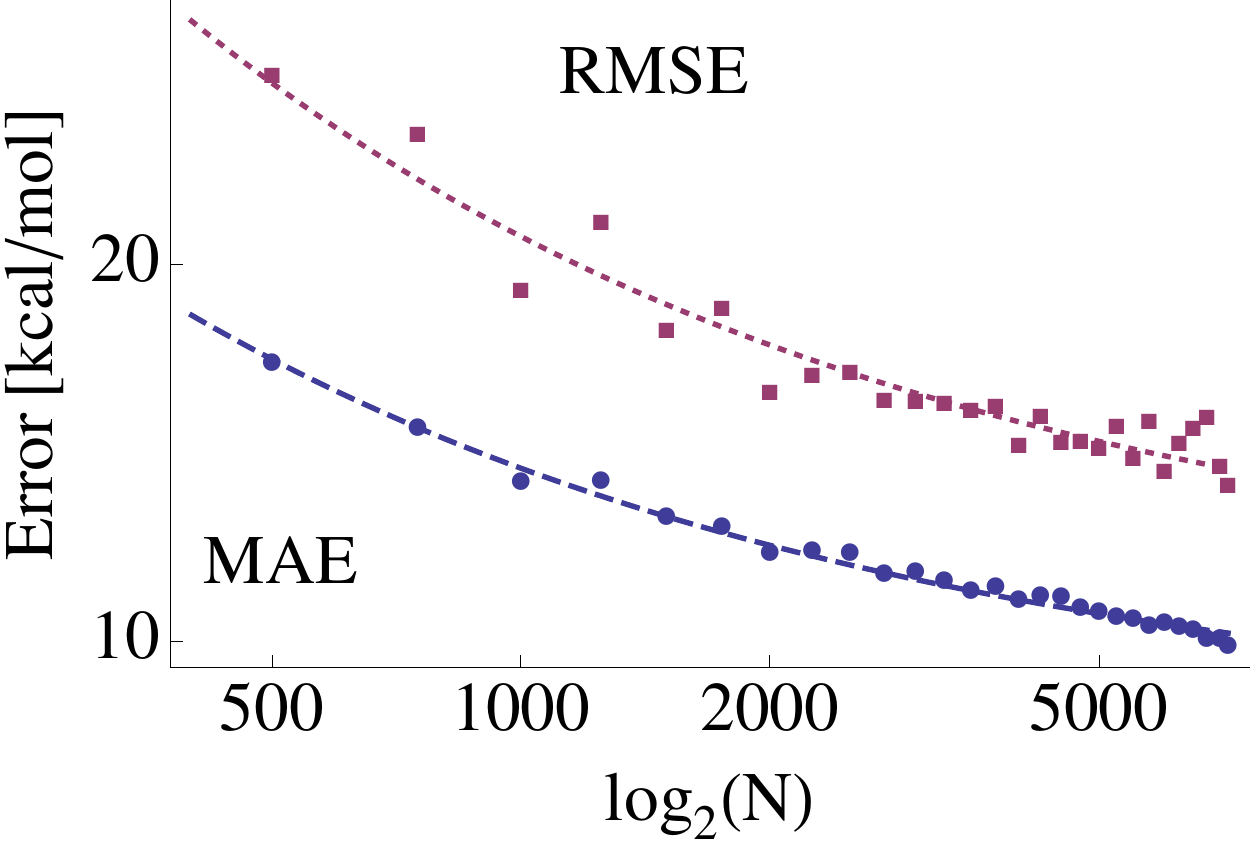}
\includegraphics[scale=0.6, angle=0]{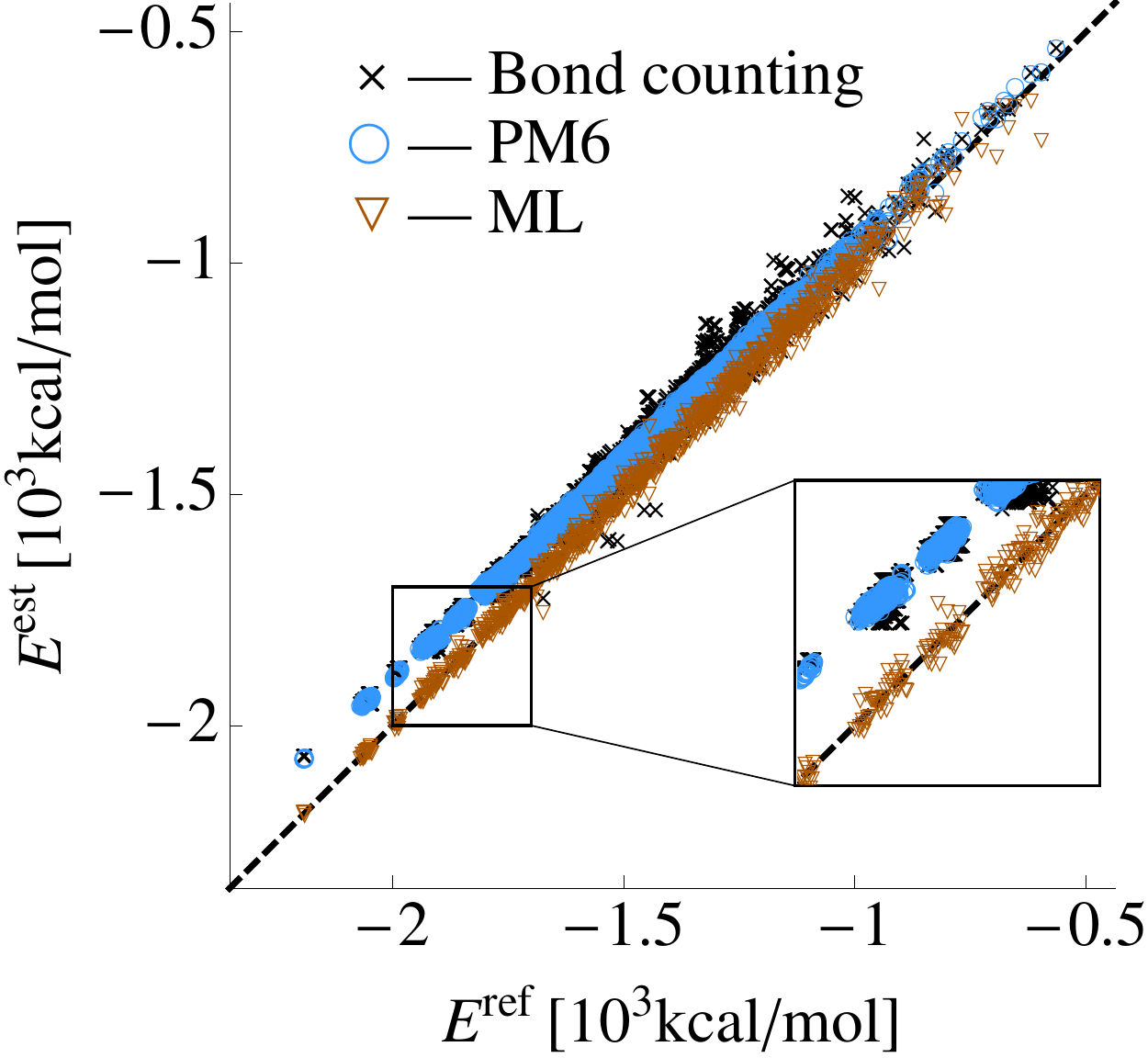}
\caption{(Color online) 
Top: Cross-validated ML errors as a function of number of molecules in training set, $N$.
Bottom: 
For $N$ = 1000, correlation of DFT-PBE0~\cite{PBE0,PBE01} results ($E^{ref}$) 
with ML (cross validated) based estimates ($E^{est}$) of atomization energies. 
Correlations for bond counting~\cite{Bondenergies} and semi-empirical quantum chemistry (PM6\cite{PM6}) are also shown.
Corresponding RMSE (root mean square error)/MAE (mean absolute error) 
for bond counting, PM6, and ML are 75.0/71.0, 75.1/73.1, 30.1/14.9 kcal/mol, respectively. 
}
\label{fig:scatter}
\end{figure}

%plot2
The dependence of the cross-validated ML performance on the number of molecules in training set, $N$, is illustrated in Fig.~\ref{fig:scatter} (top). 
When increasing~$N$ from 500 to 7000, the mean absolute error (MAE) falls off from more 
than 17 kcal/mol to less than 10 kcal/mol.
%We carried out ML for training sets 1k ($N=1000$)  and 7k ($N=7165$).
Furthermore, the width $\sigma$ of the Gaussian kernel decreases from 
$460$ to $25$  % MR: Left out standard deviation $460\pm140$ to $25\pm2.3$ to avoid having to explain this as well.
on the distance scale of Fig.~\ref{fig:distance}.
A small $\sigma$ emphasizes compound pairs for which the distance is small, 
whereas a larger $\sigma$ allows for contributions from distant pairs.
This is to be expected for increased number of training molecules.
Because of the discrete nature of chemical space 
(nuclear charges can only assume integer values), however,
we do not expect continuous coverage for $N \to \infty$,
implying that $\sigma$ will converge to a small but finite value.
The regularization hyperparameter $\lambda$ remains small throughout, consistent with the fact
that we model noise-free numerical solutions of the approximated Schr\"odinger equation.
An asymptotic fit of the form $\sim 1/\sqrt{N}$, based on statistical 
theory~\cite{htf2009,StatError_Muller1996} suggests that the MAE can be 
lowered to $\sim7.6$\,kcal/mol for $N \to \infty$. 
It is remarkable that already for the here presented, relatively small training set sizes, 
ML achieves errors of roughly one percent on the relevant scale of energies, 
clearly outperforming bond counting or semi-empirical quantum chemistry methods.
%{\em Results---}\\
%plot1
The cross-validated performance for a training set size of $N$ = 1000 is displayed in Fig.~\ref{fig:scatter} (bottom). 
There is good correlation with the DFT data. 
For comparison, corresponding correlations are shown for bond counting~\cite{Bondenergies}, 
and semi-empirical quantum chemistry (PM6~\cite{PM6}) computed with {\tt MOPAC}~\cite{MOPAC}.
While the latter two methods exhibit a systematic shift in slope, the inset highlights that 
the ML correlation accurately reproduces clustering, and slope of one.
%The two outliers of the ML model (top right in Fig.~\ref{fig:scatter}) correspond to 
%methane and acetylene, the two smallest molecules. 
%This is not surprising since the training data contains very few small molecules.

%\begin{table}[t]
%\caption{
%Correlation coefficients and out of sample prediction errors in kcal/mol for bond counting (BC)~\cite{Bondenergies},
%semi empirical (PM6) quantum chemistry~\cite{PM6}, Gaussian kernel ridge regression using the 1k and the 7k training set, 
%and transferability test (TT) using the 1k model on 6k other molecules. 
%All values rounded to two digits precision.
%} 
%\label{tab:errors}
%\begin{tabular}{lccc}
% \text{Model} & \text{RMSE} & \text{MAE} & $r^2$ \\
% \text{BC} & \text{ 74.95} & \text{ 70.96} & \text{  0.99} \\
% \text{PM6} & \text{ 75.06} & \text{ 73.08} & \text{  1.00} \\
%% \text{1k} & \text{ 20.54$\, \pm \, $1.63} & \text{ 15.62$\, \pm \, $0.98} & \text{  1.00$\, \pm \, $0.00} \\
% \text{1k} & \text{ 20.54} & \text{ 15.62} & \text{  1.00} \\
%% \text{7k} & \text{ 11.61$\, \pm \, $0.60} & \text{ 8.38$\, \pm \, $0.29} & \text{  1.00$\, \pm \, $0.00} \\
% \text{7k} & \text{ 11.61} & \text{ 8.38} & \text{  1.00} \\
% \text{TT} & \text{ 23.83} & \text{ 15.18} & \text{  0.99} 
%\end{tabular}
%\end{table}

%{\em Application---}\\
In order to assess transferability and applicability of our model to chemical compound space, 
we use a ML model trained on $N=1000$ molecules (model 1k).
The training set of model~1k contains all small molecules with 3 to 5 heavy atoms, and a randomized stratified
selection of larger compounds covering the entire energy range.
The thousand Coulomb matrices corresponding to the OpenBabel configurations were 
included as well as four additional Coulomb matrices per molecule. 
These additional matrices were scaled in order to represent the repulsive wall, the dissociative limit, 
and the energy minimum at $f=1$~\cite{RepulsiveChinese,curvesconstraints}.
All predictions are made for molecules that were {\em not} used during training of the model.

For testing the transferability, we applied the 1k model to the remaining 6k molecules. %(correspondingly augmented by the four additional matrices). 
The calculations yield errors that hardly change from the estimated performance
in the training with a MAE of 15.2~kcal/mol.
For the selected molecular subset of the seven thousand smallest molecules in the GDB database~\cite{ReymondChemicalUniverse3}, 
we therefore conclude that training on ~15\% of the molecules permits predictions of atomization 
energies for the remaining 85\% with an accuracy of roughly 15 kcal/mol. 

\begin{figure}[ht]
\centering
\includegraphics[scale=0.3, angle=90]{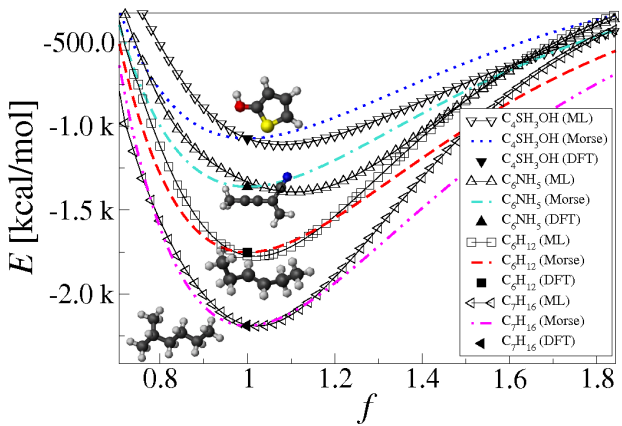}
\caption{(Color online)
Energy of atomization curves of four molecules containing single bonds and branching only (C$_7$H$_{16}$),
a double bond (C$_6$H$_{12}$), triple bonds including nitrogen (C$_6$NH$_5$), 
and a sulfur containing cycle with a hydroxy group (C$_4$SH$_3$OH).
(bottom to top in insets; black: Carbon; blue: Nitrogen; yellow: Sulfur; red: Oxygen; white: Hydrogen) 
(DFT-PBE0, Morse potential, and ML model~1k).
%Obtained parameters for Morse potentials with minimum located at f=1, fitting to E(f=2/3) = E(f=3) = 0, E(f=1) = DFT-value
%Mol A0 A1 CorrelationCoefficient DFT-value
%Hexane: 1894.11 2.08258 0.999 1895.91
%Hexyne: 1600.08 2.08258 0.999 1753.91
%Mol1: 1073.9 2.08258 0.999 1074.92
%Mol2: 1361.46 2.08258 0.999 1362.76
%Mol3=Hexene: 1752.24 2.08258 0.999 1753.91
%Mol4: 2189.89 2.08258 0.999 2191.85
} 
\label{fig:curves}
\end{figure}

For probing the applicability, we investigated whether the 1k model can also be useful beyond the equilibrium geometries. 
Specifically, we calculated the functional dependence of atomization energies on scaling Cartesian geometries by a factor, $f$. 
From the 6k molecules (not used for training) we picked four which exhibit chemical diversity. 
Specifically, these molecules contain single bonds and branching only (C$_7$H$_{16}$), 
a double bond (C$_6$H$_{12}$), triple bonds including nitrogen (C$_6$NH$_5$), and a sulfur containing cycle with a hydroxy group (C$_4$SH$_3$OH). 
The resulting ML atomization energy curves (Fig.~\ref{fig:curves}) correctly distinguish between the 
molecules, closely reproduce the DFT energy at $f = 1$,  and appear continuous and differentiable throughout relevant bonding distances. 
For comparison, corresponding Morse potential curves are also displayed.
Their well-depth and exponential factor were explicitly fitted to the molecular DFT minimum, 
as well as repulsive wall and dissociative limit~\cite{curvesconstraints}.
Albeit slightly overestimating equilibrium distance and well depth for C$_4$SH$_3$OH and C$_6$NH$_5$,
the ML model is in overall good agreement with the Morse potential curves.
One can speculate if the better performance of the ML model for the larger molecules is due to the fact that in the total set 
larger molecules are more frequent than smaller molecules. 
%or if it is because the well is deeper thereby reducing the wiggle room for the ML model.
Again, we stress the contrast that while the Morse potential curves were explicitly fitted for these
four molecules, the ML model was obtained for a training set based on one thousand {\em other} molecules.

%{\em Conclusion---}\\
We have developed a ML approach for modeling atomization energies in the chemical compound space of small organic molecules.
For larger training sets, accuracies have been achieved that are competitive with mean-field electronic structure theory, at a fraction of the computational cost.
We find good performance when making predictions for new molecules (transferability) and when predicting atomization energies beyond the equilibrium geometry.
%This is unexpected since the underlying ML statistics refrains from being stationary outside the trained domain~\cite{Sugiyamai2007,Bunau2009}.
%because  reliability of the model prediction typically deteriorates 
%This is surprising because ML generally only accounts for interpolating scenarios
%for which the statistical fine-structure of the sampled underlying probability distribution is enhanced.
Our representation of molecules as Coulomb matrices
is inspired by the nuclear repulsion term in the molecular Hamiltonian, and free atom energies. 
%Since this representation is akin to universal force-fields, 
%i.e. compositional and configurational variables are explicitly accounted for, 
%a range of possible applications is within reach, 
%including derivatives with respect to nuclear charges or atomic positions.
Future extensions of our approach might permit rational compound design applications~\cite{anatole-prl2005,anatole-jcp2009-2,CatalystSheppard2010}
as well as geometrical relaxations, chemical reactions, or 
molecular dynamics in various ensembles~\cite{tuckerman_book_SM}.
%Furthermore, our scheme could be extended towards 
%adaptive and incremental training and predictions across all of compound space, 
%in similar vein as ``Learn-on-the-fly'' for molecular force-fields~\cite{lotf2004}. 
Finally, our results suggest that the Coulomb matrix, or improvements thereof, could 
be of interest as a descriptor beyond the presented application.\\
%\cite{acknowledgmentsML}

%\section{Acknowledgments}
We are thankful for helpful discussions with
K.~Burke,
M.~Cuendet,
K.~Hansen,
J-L.~Reymond,
B.~C.~Rinderspacher,
M.~Rozgic,
M.~Scheffler,
A.~P.~Thompson,
M.~E.~Tuckerman,
S.~Varma.
All authors acknowledge support from the long program
``Navigating Chemical Compound Space for Materials and Bio Design'', IPAM, UCLA.
This research used resources of the Argonne Leadership Computing Facility at Argonne National Laboratory,
which is supported by the Office of Science of the U.S.~DOE under contract DE-AC02-06CH11357.
M.\,R.{} and K.-R.\,M.{} acknowledge partial support by DFG (MU~987/4-2) and EU (PASCAL2).
\bibliography{literatur}

\begin{thebibliography}{38}
\bibitem[{str()}]{stratification}
\bibinfo{note}{Stratification was done by sorting energies of the training
  data, grouping corresponding sorted compounds into blocks of five compounds
  each, and, for each such block, randomly assigning one compound to one
  cross-validation fold. This procedure ensures that each fold covers the whole
  energy range.}
\bibitem[{cur()}]{curvesconstraints}
\bibinfo{note}{For the repulsive wall, atomization energies for Coulomb
  matrices were set to zero at typical roots for covalent bonds,
  $f=2/3$.~\cite{RepulsiveChinese} For the minimum atomization energies, the
  finite difference derivative, $dE/df=1$, was set to zero at $f$ = 0, and
  using $df$ = 0.005. For the dissociative tail we assume zero atomization
  energies at $f=3$.}
\end{thebibliography}
\end{document}